\documentclass[aps,prl,twocolumn,groupedaddress]{revtex4-2}
\usepackage{graphicx}%
\usepackage{braket}
\newcommand{\mbi}{\affiliation{Max Born Institute for Nonlinear Optics and Short Pulse Spectroscopy, Max-Born-Straße 2a, 12489 Berlin, Germany}}%
\newcommand{\ubc}{\affiliation{Department of Chemistry and Department of Physics and Astronomy, The University of British Columbia, 2036 Main Mall, Vancouver, BC Canada V6T 1Z1}}%
\newcommand{\aremail}{\email[Email:~]{arnaud.rouzee@mbi-berlin.de}}%
\newcommand{\enemail}{\email[Email:~]{nibberin@mbi-berlin.de}}%
\newcommand{\ckemail}{\email[Email:~]{kleine@mbi-berlin.de}}%

\begin{document}


\title{Electronic State Population Dynamics upon Ultrafast Strong Field Ionization and Fragmentation of Molecular Nitrogen}
\author{Carlo Kleine}\ckemail\mbi%
\author{Marc-Oliver Winghart}\mbi%
\author{Zhuang-Yan Zhang}\mbi%
\author{Maria Richter}\mbi%
\author{Maria Ekimova}\mbi%
\author{Sebastian Eckert}\mbi%
\author{Edward R. Grant}\ubc%
\author{Marc J. J. Vrakking}\mbi%
\author{Erik T. J. Nibbering}\enemail\mbi%
\author{Arnaud~Rouz\'ee}\aremail\mbi%




\date{\today}

\begin{abstract}
Air-lasing from single ionized N$_2^+$ molecules induced by laser filamentation in air has been intensively investigated and the mechanisms responsible for lasing are currently highly debated. We use ultrafast nitrogen K-edge spectroscopy to follow the strong field ionization and fragmentation dynamics of N$_2$ upon interaction with an ultrashort 800 nm laser pulse. Using probe pulses generated by extreme high-order harmonic generation, we observe transitions indicative of the formation of the electronic ground X$^2\Sigma_{g}^{+}$, first excited A$^2\Pi_u$ and second excited B$^2\Sigma^+_u$ states of N$_2^+$ on femtosecond time scales, from which we can quantitatively determine the time-dependent electronic state population distribution dynamics of N$_2^+$. Our results show a remarkably low population of the A$^2\Pi_u$ state, and nearly equal populations of the X$^2\Sigma_{g}^{+}$ and B$^2\Sigma^+_u$ states. In addition, we observe fragmentation of N$_2^+$ into N and N$^+$ on a time scale of several tens of picoseconds that we assign to significant collisional dynamics in the plasma, resulting in dissociative excitation of N$_2^+$.
\end{abstract}


\maketitle

When intense femtosecond 800 nm laser pulses are focused in air, the combination of nonlinear self-focusing and ionization processes can lead to the formation of a filament, i.e. a narrow column of plasma formed in the wake of the propagating laser fields \cite{COUAIRON200747}. Over the last decade, the light emitted from the plasma has attracted considerable interest due to the presence of discrete lasing transitions in the spectrum of the light emitted in the forward direction \cite{luo_lasing_2003,Yao_High_2011,Liu_Self_2013,Chu_self_2013,Mitryukovskiy_Plasma_2015,xu_sub-10-fs_2015,yao_population_2016,Zhong_Vibrational_2017,Britton_Testing_2018,Mysyrowicz_Lasing_2019}. In particular, seeded \cite{yao_population_2016} and self-seeded \cite{Liu_Self_2013} cavity-free lasing at wavelengths of 391 nm and 428 nm have been observed and assigned to transitions from the second electronically excited state B$^2\Sigma^+_u$($\nu$=0) of N$_2^+$ produced by strong field ionization (SFI), to the ground electronic state X$^2\Sigma_{g}^{+}$($\nu$=0,1). Femtosecond time-resolved experiments using an additional UV seed pulse \cite{yao_population_2016,Zhong_Vibrational_2017,Britton_Testing_2018,Mysyrowicz_Lasing_2019} demonstrated that amplification at a wavelength of 391 nm can be achieved in N$_2^+$ generated by the ionizing pump laser pulse. 

As lasing requires population inversion between upper and lower electronic states, the observation of air lasing has been quite puzzling since first reported \cite{Yao_High_2011,Liu_Self_2013,Chu_self_2013}, because in the tunnel ionization regime the ionization rate is expected to decrease exponentially with the electron binding energy. In N$_2$, the ionization potentials for removal of an electron from the HOMO, HOMO-1 and HOMO-2 orbitals are 15.58 eV, 17.07 eV and 18.75 eV, respectively. Upon tunnel ionization, the population of the B$^2\Sigma^+_u$($\nu$=0) excited state of N$_2^+$ is therefore expected to be an order of magnitude smaller than the population of the X$^2\Sigma_{g}^{+}$ ground electronic state. Laser-induced inelastic recollision has been proposed as one possible mechanism that can lead to population transfer from the X$^2\Sigma_{g}^{+}$(0) to the B$^2\Sigma^+_u$($\nu$=0,1,2) excited state \cite{Mysyrowicz_PRL_2015,Britton_Testing_2018}. While recent experiments have shown a dependence of the gain on the laser ellipticity \cite{Mysyrowicz_PRL_2015}, it has been demonstrated that gain can be achieved with circularly polarized laser pulses \cite{Britton_Testing_2018}. Other studies have demonstrated the role of coherent rotational alignment dynamics in enhancing the emission at specific time-delays corresponding to a transient revival of a coherent superposition of rotational states \cite{Zhang_PhysRevX_2013,Azarm_PhysRevA_2017,Richter_Destructive_2020,Lytova_lasing_2020}. 
While rotational alignment dynamics can play an important role, the efficiency of the lasing process, in particular at a pump wavelength of 800 nm \cite{yao_population_2016,Zhong_Vibrational_2017}, cannot be explained solely by transient rotational population inversion without electronic state population inversion. 
\begin{figure}
	\centering
	\includegraphics[width=0.95\linewidth]{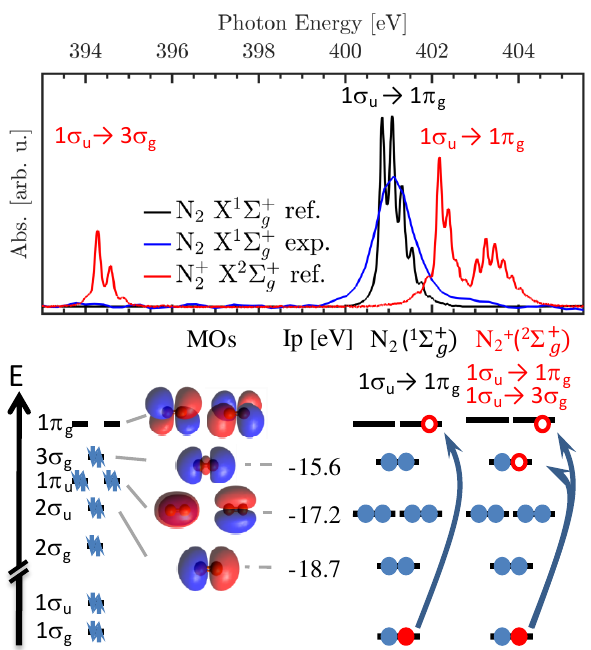}
	\caption{In the top panel, the steady-state XAS of N$_2$ measured with our high harmonic source (blue line) and at a synchrotron by \cite{Chen_PRA_1989} (black line) is shown, together with the steady-state absorption spectrum of N$_2^+$(X$^2\Sigma_{g}^{+}$) taken from \cite{lindblad_PRL_2020} (red line). In the bottom-left panel, a visualization of the level scheme of N$_2$ is provided, showing the molecular orbitals with orbital symmetry annotation and the associated ionization potentials $I_{P}$. Core excitations to the highest occupied and lowest unoccupied molecular orbitals (HOMO and LUMO) are displayed as well.} 	
	\label{fig:1}
\end{figure}

Several other mechanisms have been proposed. In some of these, population inversion between the B$^2\Sigma^+_u$-state and the X$^2\Sigma_{g}^{+}$-state is achieved through depopulation of the X$^2\Sigma_{g}^{+}$-state to the A$^2\Pi_u$-state on the trailing edge of the laser pulse \cite{xu_sub-10-fs_2015,yao_population_2016}. Alternatively, the observed lasing may be the consequence of a population inversion between the B$^2\Sigma^+_u$-state and the A$^2\Pi_u$-state \cite{Mysyrowicz_Lasing_2019,Tikhonchuk_2021_Theory}. While single-photon transitions between these states are dipole-forbidden, two-photon Raman transitions including a transition from the B$^2\Sigma^+_u$- to the X$^2\Sigma_{g}^{+}$-state are possible in the presence of a driving field that couples the X$^2\Sigma_{g}^{+}$- and the A$^2\Pi_u$-state, enabling lasing without the requirement of a population inversion between the B$^2\Sigma^+_u$- and the X$^2\Sigma_{g}^{+}$-state \cite{PhysRevA.59.3060}. This mechanism requires the possibility of a coherent laser interaction involving the B$^2\Sigma^+_u$, X$^2\Sigma_{g}^{+}$ and A$^2\Pi_u$ states under the influence of a weak post-pulse, and has been demonstrated in a previous experiment \cite{Mysyrowicz_Lasing_2019}.

In this work we present time-resolved measurements of N$_2^+$ electronic population dynamics resulting from SFI of neutral N$_2$, using femtosecond soft x-ray absorption spectroscopy (XAS) at the nitrogen K-edge. Pioneering experiments at the C and N K-edges \cite{Attar_Science_2017,
Pertot_Science_2017,Bhattacherjee_JACS_2017,
Bhattacherjee_JACS_2018,Saito_Optica_2019,Saito_Attosecond_2021,Zinchenko_science_2021},  have shown the potential of XAS for probing ultrafast molecular dynamics in isolated molecules with element- and state-sensitivity. This technique has also been applied recently to probe the SFI dynamics of CF$_3$Br \cite{Fouda_JPB_2020} and Kr \cite{Young_PRL_2006} with 100 ps time resolution using synchrotron radiation. Here, we extract the electronic population dynamics resulting from SFI of N$_2$ by XAS. Our experiment supports the formation of a population inversion between the B$^2\Sigma^+_u$ and A$^2\Pi_u$ states, which can lead to lasing by resonant Raman amplification  \cite{Mysyrowicz_Lasing_2019,Tikhonchuk_2021_Theory}.

The experimental setup is described in the supplemental material (SM). Briefly, the SFI of nitrogen induced by an intense, 800 nm laser pulse with a peak intensity of 4.5x10$^{14}$ W.cm$^{-2}$ was probed by an extreme high-order harmonic (HHG) probe pulse generated in Helium and covering a spectral range from 200-450 eV \cite{kleine_soft_2019}. The 800 nm pump and HHG probe pulses were crossed inside a gas jet containing N$_2$ molecules and the transient absorbance changes of the transmitted HHG pulse was then spectrally dispersed using a spectrometer equipped with reflective zone plate x-ray optics and an x-ray CCD detector \cite{kleine_Struc_Dyn_2021}.

The steady-state N$_2$ absorption spectrum in the absence of the pump laser pulse is shown in the inset of Fig. \ref{fig:1} (blue line) and agrees well with previously reported high-resolution spectra  \cite{Chen_PRA_1989,Ohresser_RSI_2014}. The spectrum is dominated by absorption around 401.1 eV, assigned to the transition of an electron from the $1\sigma_u(\nu=0)$ core level orbital to the $1\pi_{g}(\nu^{'})$ lowest unoccupied molecular orbital (LUMO) of N$_2$.  

\begin{figure}
	\centering
	\includegraphics[width=\linewidth]{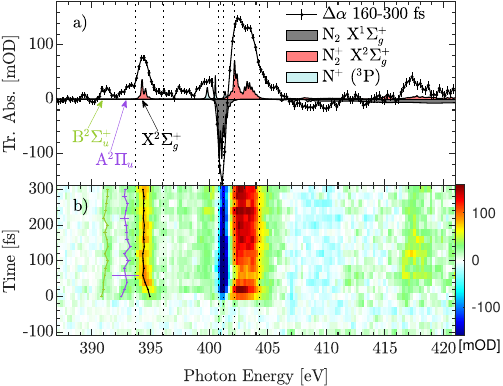}
	\caption{ a) Measured XAS integrated over time delays between 160 and 300 fs (black curve) and XAS of N$_2$(X$^1\Sigma_g$) (grey area), N$^+_2$(X$^2\Sigma_g$) (red area) and N$^+$ (blue area) reproduced from Ref. \cite{Chen_PRA_1989}, \cite{lindblad_PRL_2020} and \cite{Zeng_Single_2017}, respectively. b) Transient XAS of strong field ionized N$_2$. The central energy positions resulting from a Gaussian fit of the lowest ionic state transitions are displayed as dotted color lines (see text and Fig. \ref{fig:3}b). The spectral regions indicated by dashed lines were used to extract the temporal evolution of the absorbance at the N$_2$ 1$\sigma_u\rightarrow 1\pi_g$, N$^+_2$ (X$^2\Sigma_g$) 1$\sigma_u\rightarrow 3\sigma_g$ and N$^+_2$ (X$^2\Sigma_g$) 1$\sigma_u\rightarrow 1\pi_g$ transitions shown in Fig. \ref{fig:3}a.}	
	\label{fig:2}
\end{figure}

A series of soft x-ray spectra $I(t,E)$ were recorded by varying the time delay $t$ between the 800 nm pump pulse and the soft x-ray probe pulse in a range between -100 fs and +300 fs. In addition, scans with pump-probe delays up to 80 ps were carried out. The resulting transient N K-edge absorption spectrum is displayed in Fig. \ref{fig:2}b, together with the time-integrated transient soft x-ray spectrum averaged over the time window between 160 fs and 300 fs (see Fig. \ref{fig:2}a). A strong bleach ($\Delta$A=150 mOD, $\approx 61\pm 4\%$) of the absorbance at the $1\sigma_u\rightarrow1\pi_{g}$ core-resonance near a photon energy of 401 eV is observed upon SFI. In addition to the bleaching signal, the transient spectrum contains features in the photon energy range between 390 eV and 395 eV, two peaks around 402 eV and 403 eV and a number of weaker structures above 405 eV. Based on recent N$_2^+$ (X$^2\Sigma_{g}^{+}$) XAS measurements \cite{lindblad_PRL_2020} (see Fig. \ref{fig:1}), the absorption peak observed at 394.4 eV is attributed to the 1$\sigma_u\rightarrow$3$\sigma_g$ (HOMO) transition, which re-fills the vacancy in the molecular ion formed by SFI.  The two weak features at a photon energy of 391.2 eV and 392.9 eV, respectively, are assigned to core-transitions from the 1$\sigma_{g}$ orbital to the 2$\sigma_u$ (HOMO-2) and 1$\pi_u$ (HOMO-1) orbitals that become possible following the formation of N$_2^+$ in its second B$^2\Sigma_u^+$ and first A$^2\Pi_u$ electronically excited states. In addition, transitions from the 1$\sigma_u$ orbital to the first unoccupied 1$\pi_g$ orbital of the ground state molecular ion are responsible for the appearance of two absorption lines at a photon energy of 402 eV and 403 eV, respectively that are split by spin-orbit coupling \cite{lindblad_PRL_2020}. Within the first 300 fs following the interaction with the 800~nm laser pulse, the transient spectrum mainly contains contributions from the parent molecular ions formed by the 800 nm pump laser. However, the weak shoulder at a photon energy of 399.7 eV can be assigned to N$^+$ ions and suggests the ultrafast formation of N$^+$ by dissociative ionization \cite{Bahati_Electron_2001}.

\begin{figure}
	\centering
	\includegraphics[width=1\linewidth]{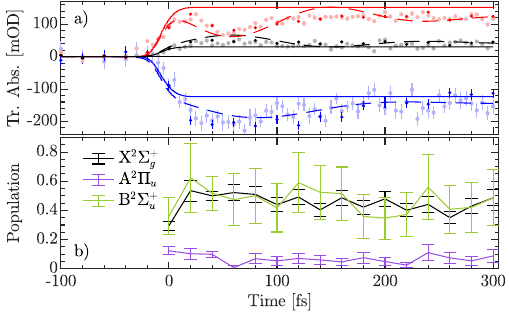}
	\caption{a) Temporal evolution of the absorbance at the N$_2$ 1$\sigma_u\rightarrow 1\pi_g$ (blue), N$^+_2$ (X$^2\Sigma_{g}^{+}$) 1$\sigma_u\rightarrow 3\sigma_g$ (black) and N$^+_2$ (X$^2\Sigma_g^{+}$) 1$\sigma_u\rightarrow 1\pi_g$ (red) transitions, spectrally integrated over the spectral regions indicated in Fig. \ref{fig:2}b by dashed lines. Simulations of the transient signals accounting for ionization only (solid lines) and ionization and alignment (dashed lines) are also shown in panel a). b) Ultrafast population dynamics of the X$^2\Sigma_{g}^+$, A$^2\Pi_{u}$ and B$^2\Sigma_u^+$ states of N$_2^+$ retrieved using Eq. \ref{eq:1}.} 
	\label{fig:3}
\end{figure}

The temporal evolution of the transient absorbance at the N$_2$ 1$\sigma_u\rightarrow 1\pi_g$, N$^+_2$ 1$\sigma_u\rightarrow 3\sigma_g$ and N$^+_2$ 1$\sigma_u\rightarrow 1\pi_g$ resonances near 401 eV, 395 eV and 403 eV, respectively, are shown in the Fig. \ref{fig:3}a. Within the first 25 fs following the temporal overlap, SFI leads to a decrease of the absorbance at the N$_2$ 1$\sigma_u\rightarrow 1\pi_g$ core-resonance that is accompanied by an increase of the absorbance at both the N$^+_2$ 1$\sigma_u\rightarrow 3\sigma_g$ and 1$\sigma_u\rightarrow 1\pi_g$ resonances. More interestingly, a strong modulation is observed in the N$^+_2$ 1$\sigma_u\rightarrow 1\pi_g$ (LUMO) transient signal, with a minimum observed at a time delay of $\approx$ 50 fs. A similar effect has recently been observed in SFI of NO molecules \cite{Saito_Optica_2019} and was attributed to alignment dynamics of the molecule on the leading edge of the 800 nm pump laser pulse. In moderately intense laser fields (typically below 5$\cdot$10$^{13}$ W.cm$^{-2}$), the induced dipole moment from the interaction of the molecular polarizability with the pulse can lead to a rotational wavepacket that rephases periodically after the laser pulse has ended, leading to a transient alignment of the molecular axis along the laser polarization axis \cite{Vrakking_PRL_2001,Peronne_PRL_2003,Karamatskos_NatComm_2019} that is delayed with respect to the maximum of the pump laser pulse.  Since the $\sigma\rightarrow\pi$ transition dipole moment is perpendicular to the molecular axis, we expect a transient decrease of the absorption when the molecules are aligned along the laser polarization axis.
To validate this assumption, simulations were performed in which focal-averaged ionization dynamics were calculated based on the Ammosov-Delone-Krainov (ADK) model \cite{tong_theory_2002,zhao_determination_2010}. In addition, the laser-induced field-free alignment dynamics was computed by solving the time-dependent Schr\"odinger equation in order to evaluate the expectation value $\braket{\cos^2\theta(t)}$, with $\theta$ the angle between the laser polarization axis and the molecular axis. The time-dependent absorption at the three resonances shown in Fig.~\ref{fig:3}a was then evaluated assuming a $\braket{\sin^2\theta}=(1-\braket{\cos^2\theta}$) dependence for the N$_2$ 1$\sigma_u \rightarrow 1\pi_g$ and N$_2^+$ 1$\sigma_u\rightarrow 1\pi_g$ core-transitions and a $\braket{\cos^2\theta}$ dependence for the 1$\sigma_u\rightarrow 3\sigma_g$ core-transition (see SM). The result of the model is displayed in Fig.~\ref{fig:3}a (dashed lines). We note that for this comparison, the amplitude of the absorbance change in our model was scaled to the measurements. Our model reproduces the measured transient absorbance at the three core-resonances, including the minimum in the $1\sigma_u\rightarrow 1\pi_g$ absorbance observed near 50 fs, confirming the role of molecular alignment in the observed dynamics. 

The electronic state selectivity of soft x-ray absorption spectroscopy allows to derive the time-dependent electronic state population distribution of the N$_2^+$ ions when the relative electronic oscillator strengths $f_j$ for a 1s core electron to re-fill the electronic states of N$_2^+$ are known. They can be experimentally obtained by monitoring the x-ray emission corresponding to transitions from the three valence electronic states to the 1s core-hole state of N$_2^+$ formed by removal of a 1s electron from N$_2$ by x-ray ionization above the N K-edge. From reported x-ray emission spectra \cite{Glans_Selectively_1996}, the relative electronic transition oscillator strengths are 0.8, 1.0 and 0.15 for the X$^2\Sigma^+_g$(3$\sigma_g^{-1}$)$\leftrightarrow$N$_2^+$($\sigma_u^{-1}$), A$^2\Pi_u$(1$\pi_u^{-1}$)$\leftrightarrow$N$_2^+$($\sigma_g^{-1}$) and B$^2\Sigma^+_u$(2$\sigma_u^{-1}$)$\leftrightarrow$N$_2^+$($\sigma_g^{-1}$) transitions, respectively. Using these values, the relative $N^+_2$ state-populations after the pump pulse has ended are obtained by fitting the corresponding absorption lines in the measured transient soft x-ray spectrum with a Gaussian function. The relative populations are then given by:
\begin{equation}
	\rho_{ij}=\frac{A{ij}\cdot w_{ij}}{f_{j}}/N,\, N=\sum_j \frac{\overline{A}{j}\cdot\overline{w}_{j}}{f_{j}},
	\label{eq:1}
\end{equation}
with $A_{ij}$ and $w_{ij}$ the amplitude and the width of the Gaussian fits for each transition $j$ at each time delay $i$, and $N$ a normalization factor retrieved from the measurements at delays between 160 fs to 300 fs. The result of this procedure is shown in Fig. \ref{fig:3}b. Surprisingly, after SFI the populations of the X$^2\Sigma^+_{g}$ and B$^2\Sigma^+_u$ states of $0.47\pm0.02$ and $0.46\pm0.07$, respectively, are almost identical while the population of the A$^2\Pi_{u}$ state remains low ($0.07\pm0.01$). We note that during the interaction with the ionizing laser pulse, a shift of the central energy of the fitted Gaussians can be observed (dotted lines in Fig. \ref{fig:2}b) and assigned to an AC Stark effect \cite{Chini_PRL_2012,Ott_Science_2013,Warrick_JPCA_2016,Drescher_JPCL_2019}. 

To the best of our knowledge, our experimental results provide the first direct indication that under ultrafast SFI conditions the N$^+_2$ B$^2\Sigma_u^+$ excited state can reach a population with a similar magnitude as that of the N$^+_2$ X$^2\Sigma_{g}^+$ electronic ground state. In addition, the population of the N$^+_2$ A$^2\Pi_{u}$ state remains small after the SFI process. As such, our results are at odds with the predictions of tunnel ionization models, in which the population of excited cationic states decreases exponentially with the electron binding energy. In addition, our results contrast with theoretical predictions \cite{yao_population_2016}, performed by solving the time-dependent Schrödinger equation for conditions where lasing has been observed in filamentation experiments, i.e. at a lower intensity compared to the one used in our study. As a main result, these calculations have shown that the N$_2^+$ state-population can be redistributed during the laser pulse, and that a depopulation of the ground electronic state of the molecular ion to the A$^2\Pi_{u}$ state can occur through efficient quasi-resonant one-photon coupling at 800 nm \cite{yao_population_2016}. In turn, this redistribution allows for population inversion between the B$^2\Sigma_u^+$ and X$^2\Sigma_{g}^+$ states, while the A$^2\Pi_{u}$ state plays the role of a population reservoir. Our measurement performed at a higher intensity suggest that there is no population accumulation in the A$^2\Pi_{u}$. In contrast, our results are consistent with lasing without electronic state population inversion as proposed by Mysyrowicz et al. \cite{Mysyrowicz_Lasing_2019,Tikhonchuk_2021_Theory} that can be achieved by Raman amplification when the B$^2\Sigma_u^+$ state population is smaller than that of the X$^2\Sigma_{g}^+$ state, but larger than that of the A$^2\Pi_{u}$ state. However, our results cannot directly prove the validity of this mechanism. 
\begin{figure}
	\centering
	\includegraphics[width=\linewidth]{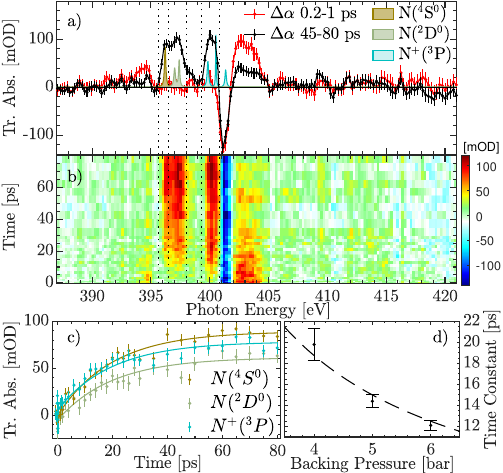}
	\caption{a) Transient soft x-ray spectra, integrated for short (0.2-1.0 ps) and long (45-80 ps) delays, revealing the disappearance of the N$_2^+$ contributions and the appearance of N and N$^+$ contributions. Reference spectra (colored area) have been taken from an R-matrix calculation reported in Ref. \cite{Zeng_Single_2017} and are energy shifted to match the experimental values reported in \cite{santanna_k_2011,gharaibeh_k-shell_2012}. b) Transient XAS in the time window between -0.6 ps and 80 ps. c) Temporal evolution of N and N$^+$ absorption signals (dots) spectrally integrated over the spectral regions indicated by the dashed lines in panel a) corresponding to 1s$\rightarrow$2p transitions in the $^4S^0$, $^2D^0$ and $^3P$ states of N and N$^{+}$. The kinetic fits (solid lines) have been derived by singular value decomposition using a single time constant. d) N$_2^+$ dissociation time as a function of the N$_2$ backing pressure (dots) and corresponding fit (dashed line) using an inverse function $1/p$.}		
	\label{fig:4}
\end{figure}

Additional experiments were performed in which soft x-ray absorption spectra were recorded for pump-probe time delays ranging from -0.6 ps to 80 ps. The corresponding transient N K-edge absorption spectrum is shown in Fig. \ref{fig:4}b. In the photon energy region between 402 eV and 404 eV in Fig. \ref{fig:4}b, the absorption associated with transitions occurring in the ground state molecular ion rapidly decays, indicative that under our experimental conditions the N$_2^+$ ions do not survive on longer picosecond time scales. Instead, a major fraction is converted into atomic N and N$^+$, as evidenced by new transitions observed around 396 eV, 397 eV and 399 eV that are assigned to core level transitions in N and N$^+$ (see Fig. \ref{fig:4}a). Using a singular value decomposition procedure, the rise-time of the atomic N and N$^+$ contributions, and the concomitant decay time of the N$_2^+$ molecular ion components, can be extracted from our measurements (Fig. \ref{fig:4}c). We deduce a time constant of $\tau=$19.8~$\pm$~1.5~ps for an N$_2$ backing pressure of 4 bar (Fig. \ref{fig:4}c). Interestingly, this time constant strongly depends on the backing pressure used as displayed in Fig. \ref{fig:4}d, which shows that the time constant decreases when the pressure increases. We argue that this observation is caused by collisional processes in the plasma that is formed at the focus of the ionizing laser pulse. Dissociative electron-ion recombination is expected to be of minor importance in our experimental conditions. Instead, we expect that the density of energetic electrons is large enough to lead to significant collisional dynamics in the plasma, resulting in dissociative excitation of N$_2^+$ ions to form both neutral and singly charged atomic nitrogen. The kinetic energy distribution $\rho (E)$ of electrons that result from SFI mainly consists of a broad plateau that extends to a kinetic energy of 2$U_p$ \cite{meckel_signatures_2014}, with $U_p$ the ponderomotive energy of the laser pulse given by $U_p$(eV)=9.33$\cdot I$($10^{14}$W.cm$^{-2}$)$\cdot \lambda^2$($\mu m$)=29 eV for a peak intensity $I$ of $\approx$ 4.5x10$^{14}$ W/cm$^2$. The cross-section for dissociative excitation for electrons as a function of the kinetic energy has been previously characterized \cite{peterson_dissociative_1998}. Knowing the gas density and the cross-section for dissociative excitation $\sigma (E)$, and assuming that 60 $\%$ of molecules are ionized in the focal volume, the rate $k$ for dissociative excitation can be estimated using:
\begin{equation}
k=n_e\int \sqrt{\frac{2E}{m_e}}\rho(E)\sigma(E)dE
\end{equation}
with $n_e$ the electron density. Using this expression, we can estimate that dissociative excitation leading to atomic fragments will occur with a rise-time $\tau=1/(k\,ln(2))$ of $\approx$ 28 ps, which is relatively close to the observed value of $\tau=19.8\pm1.5$ ps for a backing pressure or 4 bar (see SM).

In conclusion, we have investigated the SFI dynamics of N$_2$ by intense 800 nm laser pulses using femtosecond soft x-ray absorption spectroscopy. As a main result, we characterized the electronic population distribution of N$_2^+$ ions that are formed. We found that the A$^2\Pi_{u}$-state population is small compared to the population in the X$^2\Sigma_{g}^{+}$ ground and B$^2\Sigma_u^+$ second excited states, which are almost equally populated. Our results invalidate the role of the A$^2\Pi_{u}$ state as a reservoir to achieve population inversion between the B$^2\Sigma_u^+$ and X$^2\Sigma_{g}^{+}$ states, and thus rule out proposed models of N$_2$ lasing that have been based on this \cite{yao_population_2016}. However, our results are compatible with proposed models \cite{Mysyrowicz_Lasing_2019,Tikhonchuk_2021_Theory} for air lasing based on a V-type level scheme involving the three first electronic states of N$_2^+$ or models for air lasing based on rotation-induced population inversion between the B$^2\Sigma_{u}^{+}$ and X$^2\Sigma_{g}^{+}$ states \cite{Zhong_Vibrational_2017,Richter_Destructive_2020,Lytova_lasing_2020}.

\begin{acknowledgments}
This work has been supported by the European Research Council (ERC) under the European Union’s Horizon 2020 research and innovation programme (ERC Grant Agreement N$^\circ$ 788704; E.T.J.N.) and the ITN network SMART-X (860553). ERG gratefully acknowledges support from AFOSR FA9550-17-1-0342. We gratefully acknowledge fruitful discussions with M. Ivanov, F. Morales, S. Patchkovskii and A. Husakou.
\end{acknowledgments}


\renewcommand{\bibsection}{\section*{References}}
\bibliography{PRL_Rouzee_2022_ref}

\end{document}